\input{aipcheck}

\documentclass[
    ,final            
  ]
  {aipproc}

\layoutstyle{6x9}

\begin{document}

\title{Phase Transitions in Finite Systems using Information Theory}


\author{Philippe CHOMAZ}{
  address={GANIL (DSM-CEA/IN2P3-CNRS), BP 55027, F-14076 Caen c\'edex 5, France}
}

\author{Francesca GULMINELLI}{
  address={LPC (IN2P3-CNRS/Ensicaen et Universit\'e), F-14076 Caen c\'edex, France}
}


\begin{abstract}
In this paper, we present the issues we consider as 
essential as far as the statistical mechanics 
of finite systems is concerned. In particular, we emphasis our present understanding of 
phase transitions in the framework of information theory. 
	Information theory provides a thermodynamically-consistent treatment of 
finite, open, transient and expanding systems
which are difficult problems in approaches using standard statistical ensembles. 
As an example,  we analyze is the problem of boundary conditions, which in the framework of 
information theory must also be treated statistically. 
We recall that out of the thermodynamical limit the different ensembles are not equivalent and in particular they may lead to dramatically different equation of states, 
 in the region of a first order phase transition.  
	We recall the recent progresses achieved in the understanding of first-order phase transition in finite systems:  the equivalence between the Yang-Lee theorem and the occurrence of bimodalities in the intensive ensemble and the presence of inverted curvatures of the thermodynamic potential 
of the associated extensive ensemble. We come back to the concept of order parameters and to the role of constraints on order parameters in order to predict the expected signature of  first-order phase transition: in absence of any constraint (intensive ensemble) bimobality of the event distribution is expected while an inverted curvature of the thermodynamic potential is expected at a fixe value of the order parameter (extensive ensemble) in between the phases (coexistence zone). 
	We stress that this discussion is not restricted to the possible occurrence of negative specific heat, 
but can also include negative compressibility's and negative susceptibilities,
and in fact any curvature anomaly of the thermodynamic potential.
\end{abstract}

\maketitle


\section{Introduction: Unusual Mesoscopic Worlds}
\label{intro}
Everybody knows that when a liquid is heated, its temperature increases
until the moment when it starts to boil. The increase in temperature then
stops, all heat being used to transform the liquid into vapor. What is the
microscopic origin of such a strange behavior? Does a liquid drop containing
only few molecules behave the same? Recent experimental and theoretical
developments seem to indicate that at the elementary level of very
small systems, this anomaly appears in an even more astonishing way: during
the change of state - for example from liquid to gas - the system cools
whereas it is heated, i.e. its temperature decreases while its energy
increases. This phenomenon is only one example of the fact that small 
systems when heated or compressed do not behave like the macroscopic 
systems we are used to. This paper presents a review of key issues about
the thermodynamics of small systems in particular in presence of phase 
transition.

In many different fields of physics,
finite systems properties, non extensive thermodynamics, and phase transitions out of the thermodynamic limit are strongly debated issues(see for example \cite{Dauxois3}). This may be the case of non saturating forces such as the gravitational \cite{Lynden-Bell2,lynden-bell,ruffo,tatekawa} or the Coulombic forces. 
The system may be too small, as in the case of clusters and nuclei \cite{Schmidt,Agostino,Farizon,Melby}. The physics of finite systems is even more complicated since often they are not only small but also open and transient. 
This implies that the various concepts of thermodynamics and statistical mechanics
\cite{landau,huang,tolman,rasetti} 
have to be completed and revisited \cite{Dauxois3,jaynes,gross,balian,Hill,tsallis,chomaz-here}. 

A consistent framework to address those issues is the information theory approach to statistical 
mechanics\cite{jaynes,balian}. This formalism 
allows to address in a consistent way the statistical mechanics of open systems
evolving in time, independent of their interaction range and number of constituents. 
After a short summary of the statistical physics concepts, 
we will first address the essential question of boundary conditions, 
which cannot be avoided when dealing with finite systems with continuum states.
Recalling that the exact knowledge of a boundary requires 
an infinite information, we will show that a consistent treatment of unbound systems
can still be achieved if boundary conditions are treated statistically,
leading to new statistical ensembles.
Then we will recall that for finite systems the different ensembles 
are not equivalent
\cite{tsallis,barre2,leyvraz,Gulminelli,Costeniuc} 
In particular, two ensembles which put different constraints on the fluctuations 
of the order parameter lead to qualitatively different 
equations of state close to a first order phase transition\cite{Gulminelli}.

As an example, when energy is the order parameter, 
the microcanonical (at fixed energy) heat capacity diverges to become negative while the 
canonical (at fixed temperature) one remains always positive and finite
\cite{gauss,Thirring,Huller,Ellis,Dauxois,Barre,Ispolatov,Kastner,Gross3,Huller2,Pleimling,Chomaz02,Chavanis}. 
If the magnetization is the order parameter, it is the magnetic susceptibility,
or if the volume (or density) is the order parameter, it is the compressibility, 
or if the number of particle is the order parameter, it is the chemical susceptibility 
which are expected to present a negative branch in between two divergences in the fixed 
order parameter ensemble while in the ensemble in which the order parameter has 
only a mean value constrained by a Lagrange intensive parameter (magnetic field or pressure or
chemical potential)
should remain positive. 
This difference between ensembles can be of primordial importance for microscopic and 
mesoscopic systems 
undergoing a phase transition. Such systems are now studied in many fields of physics, from 
Bose condensates \cite{bose,bose2} to the quark-gluon plasma \cite{qgp,high:energy}, 
from cluster melting \cite{Schmidt,clusters} to nuclear fragmentation \cite{Agostino}. 
Moreover, such inequivalences may survive at the infinite size limit 
for systems involving long range forces such as self-gravitating 
objects\cite{lynden-bell,ruffo,tatekawa}.

Then, going deeper insight the formalism, 
we will summarize the mathematical equivalence between the 
Yang-Lee approach \cite{Yang} through the zeroes of the partition sum
in the space of complex intensive parameters associated with an order parameter, 
the bimodality of the order parameter distribution in this intensive ensemble,
and the anomalous (inverted) curvature of the thermodynamic potential 
of the ensemble where the order parameter is fixed\cite{topology,Lee,Chomaz3}. 
The best documented example in the literature is the bimodality of the 
canonical energy distribution and its equivalence to negative microcanonical 
heat capacity\cite{binder,labastie,europhys}.  
As far as the time evolution problem is concerned, 
we stress the need to take into account time odd constraint in the 
statistical picture. Finally, we conclude about the important challenges related
finite-systems thermodynamics.

\section{Statistical Mechanics Description of Finite System}

When discussing thermodynamics, people often implicitly  refer to macroscopic 
systems. Indeed, when it is possible to take the thermodynamical limit of 
infinite systems the statistical physics simplifies since all statistical ensembles 
are equivalent. However, when this limit is not taken because it does not exist
for the considered system (e.g. self-gravitating systems) or because the 
considered system is simply finite, one may think that the thermodynamics
concept are becoming vague or even ill-defined.  Are concepts like 
equilibrium, temperature, pressure etc. applicable to objects
as tiny as nuclei. How large must be a system for a temperature to be defined?
These questions originates from a confusion between a possible uncertainty 
about the statistical ensemble corresponding to the studied system and a fundamental
problem on thermodynamical concepts. Only the former is a real issue. Indeed, 
when a statistical ensemble is defined all thermodynamical quantities are clearly 
mathematically defined. At the basis of this discussion, the concept of equilibrium 
is also well defined since in the Gibbs spirit it is nothing but the ensemble 
of events maximizing the entropy under the considered constraints.  Then, a concept
like the temperature is well define the problem being that it is not the same definition 
in the various ensemble. then the question is what is the physical meaning of 
thermodynamic quantities - say, temperature - evaluated through different ensembles?
is there a "`correct"' ensemble to be used? 

This discussion becomes even stronger when phase transitions are concerned.
Indeed, when the thermodynamical limit can be taken phase transitions are well 
defined in all text book as non-analytical properties of thermodynamical potentials. 
However, this anomalous behavior being generated when taking  the limit of infinite 
  systems this definition of phase transition cannot be operational for finite 
  systems. However, we all know finite systems can change
state or shape, a typical example being the case of isomerization; how many
degrees of freedom do we need in order to call this change of 
state a phase transition?

To start answering those questions about the thermodynamics concepts 
applied to finite systems we can first look at simple systems. 
Let us consider a system that can exist in two single microstates of different energy  
(a single spin in a magnetic field, a two-level atom in a bath of radiation...)
The system being much smaller than its environment, 
let us consider the case for which
the interaction between system 
and environment can be neglected and we have no reason to believe that the environment
will be in any specific state. Then the distribution of the system microstates
is simply given by the number of states of the environment

\begin{eqnarray}
p^{(n)}&=&W(E_t-e_n)/(W(E_t-e_1)+W(E_t-e_2))\nonumber \\
&\propto& exp \left ( S(E_t-e_n) \right )
\label{lipkin}
\end{eqnarray} 

where $E_t$ is the total energy (system + environment) and 
$S=\log W$ is the (microcanonical) entropy associated to the environment.
Since $e_n \ll E_t$, a Taylor expansion of the entropy gives

\begin{equation}
S(E_t-e_n) \approx S(E_t)-e_n\frac{\partial S}{\partial E} (E_t)
\; \; ; \; \;
p^{(n)}\propto exp \left ( - \beta e_n \right)
\end{equation}
where we have introduce $\beta={\partial S}{\partial E}$, the temperature of the environment.  

This very simple textbook exercise gives us a number of interesting information:

\begin{itemize}
\item thermodynamic concepts like temperature can be defined for systems having an 
arbitrary number of degrees of freedom (the minimum being 
2 levels)
\item 
Boltzmann-Gibbs statistics naturally emerges as soon as we observe
a limited information constructed from a reduced number of degrees of freedom
\end{itemize}

If we now take into account a slightly more complicated system with 
energy states associated to a degeneracy $w(e)$, the energy distribution 
will be modified to

\begin{equation}  
p(e)=\frac{w(e)W(E_t-e)}{\sum_n w(e_n)W(E_t-e_n)}
\approx \frac{w(e)exp\left ( - \beta e \right)}{Z_\beta}
\label{lipkin2}
\end{equation}
where the canonical approximation is still correct if the system is associated 
to a much smaller number of degrees of freedom than its environment.
Eq.(\ref{lipkin2}) gives for instance the energy distribution of a thermometer 
loosely coupled to an otherwise isolated system.
Temperature is defined as the response of the thermometer 
in the most probable energy state $\overline{e}$;
if we maximize the distribution (\ref{lipkin2}) we get,
assuming that energy can be treated as a continuous variable

\begin{equation}
\frac{\partial \log W}{\partial E}|_{E_t-\overline{e}}=
\frac{\partial \log w}{\partial E}|_{\overline{e}}
\end{equation} 
We then learn that the quantity shared at the most probable energy partition is the microcanonical
temperature. 
This shows that there is no
ambiguity in the definition of temperature (and any other thermodynamic quantity) when dealing 
with small systems.

It is important to note that eq.(\ref{lipkin2}) is not limited to the observation of energy, but 
can apply to the distribution of any generic observable $A=\langle \hat{A} \rangle$. We can then expect 
that canonical-like ensembles (i.e. ensembles where distributions are given by Boltzmann factors)
will arise each time that we are isolating a small number of degrees of freedom from a more complex
system. 
More generally, we will recall in the next section that a statistical description is in order 
each time that the system is complex enough to have a large number of microstates associated to a 
given set of relevant observables. If the relevant observables are recognized, equilibrium is therefore a very generic concept of minimum information theory. The proper statistical ensemble i.e. the relevant observables depend on the dynamics of the considered  system and the way it is prepared.

\section{Information Theory}\label{sec:Stat}

Information theory is the general framework which provides the foundation of statistical mechanics. 
It should be noticed that it also gives the framework for the generalization of the Density Functional Theory (DFT) \cite{DFT}.  
It leads to a consistent treatment of the thermodynamics of finite systems both in the classical and in the quantal world\cite{balian,jaynes}. 
Let us summarize here the essential ingredients. 
We will use here quantum mechanics notations. Classical approaches can immediately be defined as a classical limit of the presented results.
Statistical physics treats statistical ensembles of possible solutions for
the considered physical system. 

\subsection{Liouville Space}\label{ssec:Liou1}

Such a ''macro-state'' can be 
represented by its density matrix

 \begin{equation}
 \hat{D}=\sum_{\left( n\right) }\left| \Psi ^{\left( n\right) }\right\rangle\;
 p^{\left( n\right) }\;
 \left\langle \Psi ^{\left( n\right) }\right|,
 \end{equation}
 where the states ("micro-states", or "partitions", or "replicas", or simply "events") 
 $\left| \Psi ^{(n)}\right\rangle $ pertain to the considered Fock or Hilbert space. 
 $p^{\left( n\right)}$ is the occurrence probability of the event $\left| \Psi^{(n)}\right\rangle $. 
 The result of the measurement of an observable $\hat{A }\ $is 

\begin{equation}
<\hat{A}>_{\hat{D}}=\mathrm{Tr}\hat{A}\hat{D},
\label{Eq:trAD}
\end{equation}
where $\mathrm{Tr}$ means the trace over the quantum Fock or Hilbert space of states 
$\{ \left| \Psi \right\rangle \} $. 
\begin{figure}[tbp]
\resizebox{0.75\columnwidth}{!}
{\ \includegraphics{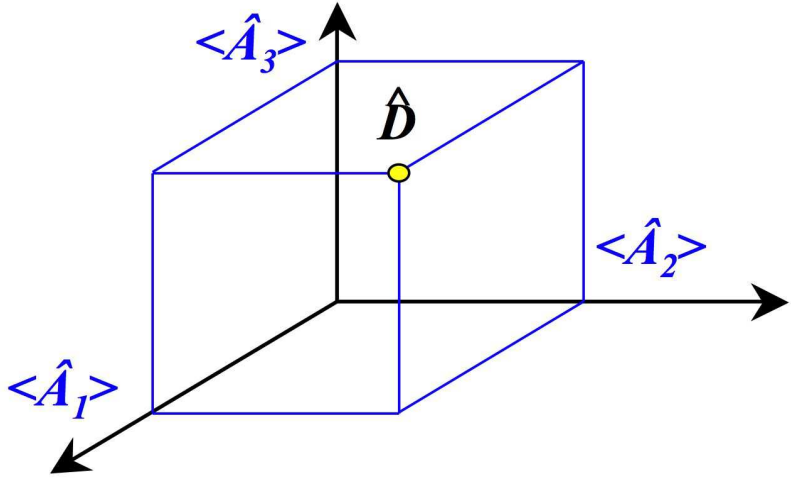} }
\caption{Illustration of the Liouville space of density matrices: an observation 
$\langle \hat A_i \rangle$ is a projection 
of $\hat D$ on the axis associated with the corresponding observable $\hat A_i$.}
\label{fig:2}
\end{figure}
In the space of Hermitian matrices, 
the trace provides a scalar product \cite{Reinhardt,balian2}

 \begin{equation}<<\hat{A}||\hat{D}>>=\mathrm{Tr}\hat{A}\hat{D}.  \label{EQ:scalarAD}
\end{equation}
It is then possible to define an orthonormal basis of Hermitian operators $\{ \hat{O}_{l}\} $
in the observables space, and to interpret the measurement $<\hat{O}_{l}>_{\hat{D}}$ 
as a coordinate of the density matrix $\hat{D}$ (see fig.\ref{fig:2}). 
The size of the observables space is the square 
of the dimension of the Hilbert or Fock space, which are in general infinite; 
therefore in order to describe the system, one is forced to consider a reduced set of 
(collective) observables 
$\{ \hat{A}_{\ell }\} $ which are supposed to contained the relevant information. 

\subsection{Maximum Entropy State and Generilized Gibbs Equilibrium}\label{ssec:Liou}

The Gibbs formulation of statistical mechanics can then be derived if the least 
biased''macro-state'' is assumed to be given by the maximization of  the entropy
\footnote{In this article we implicitly use units such that the Boltzmann constant $k=1$.}

\begin{equation}
S[\hat{D}]=-\mathrm{Tr}\hat{D}\log \hat{D},  
\label{EQ:S}
\end{equation}
which is nothing but the opposite of the Shannon information\cite{balian,jaynes}.
It is important to notice that this formalism can be generalized introducing 
alternative entropies such as the Tsallis entropy.  
Eq.(\ref{EQ:S}) is a definition of entropy valid for any density matrix
which coincides with the standard thermodynamic entropy only after maximization, see eq.
(\ref{EQ:Legen}) below.

If the system is characterized by $L$ (relevant) observables (or ''extensive'' variables
\footnote{In this paper the word ''extensive'' is used in the general sense of resulting from an 
observation, i.e. the $<\hat{A}_{\ell }>,$ and not in the restricted sense of additive variable. 
Intensive variables are conjugate to extensive variables i.e. Lagrange multipliers $\lambda _{\ell }$ imposing the average value of the associated extensive variable.}), 
$\mathbf{\hat{A}}=\{ \hat{A}_{\ell }\}$, 
known in average $<\hat{A}_{\ell }>=\mathrm{Tr}\hat{D} \hat{A}_{\ell }$, 
the variation of the density matrix is not free. The standard  way to solve this maximization 
of the entropy under constraints is to maximize with no constraints the constrained entropy

\begin{equation}
S^{\prime }=S-\sum_{\ell }\lambda _{\ell }<\hat{A}_{\ell }>, 
\end{equation}
where the $\mathbf{\lambda }=\{\lambda _{\ell }\}$ 
are $L$ Lagrange multipliers associated with the $L$ constraints $<\hat{A}_{\ell }>$.

A maximization of the entropy under constraints
gives a prediction for the minimum biased density matrix (or "`event distribution"')
which can be viewed as a generalization of 
Gibbs equilibrium:

\begin{equation}
\hat{D}_{\mathbf{\lambda }}=\frac{1}{Z_{\mathbf{\lambda }}}\exp -\mathbf{\lambda .\hat{A}},  
\label{EQ:D}
\end{equation}
where $\mathbf{\lambda .\hat{A}=}\sum_{\ell =1}^{L}\lambda _{\ell }\hat{A}_{\ell }$ and where $Z_{\mathbf{\lambda }}$ is the associated partition sum insuring the normalization of $\hat{D}_{\mathbf{\lambda }}$: 

\begin{equation}
Z_{\mathbf{\lambda }}=\mathrm{Tr}\exp -\mathbf{\ \lambda .\hat{A}} 
\end{equation}
Using this definition, we can compute the associated equations of state (EoS): 

\begin{equation}
<\hat{A}_{\ell }>=\partial _{\lambda _{\ell }}\log Z_{\mathbf{\lambda }}.
\label{EQ:EOS}
\end{equation}
The entropy associated with $\hat{D}_{\mathbf{\lambda }}$ is: 

\begin{equation}
S[\hat{D}_{\mathbf{\lambda }}]=\log Z_{\mathbf{\lambda }}
+\sum_{\ell }\lambda _{\ell }<\hat{A}_{\ell }>,  
\label{EQ:Legen}
\end{equation}
which has the structure of a Legendre transform between the entropy and the thermodynamic potential.

\subsection{Ergodicity and Microcanonical Ensemble }\label{ssec:Ergo}

To interpret the Gibbs ensemble as resulting from the contact with a reservoir or to guarantee the stationarity of eq.(\ref{EQ:D}), it is often assumed that the observables $\hat{A}_{\ell }$ are conserved quantities such as the energy $\hat{H}$, the particle (or charge) numbers $\hat{N}_{i}$
or the angular momentum $\hat{L}$\cite{gross-here}. 
However, there is no formal reason to limit the state variables to constants of motion. Even more, 
the introduction of not conserved quantities might be a way to take into account some non ergodic aspects.
Indeed, an additional constraint reduces the entropy, limiting the populated phase space or modifying the event distribution. 
This point will be developed at length in the next sections.

It should be noticed that microcanonical thermodynamics also corresponds to a maximization of the entropy (\ref{EQ:S}) in a fixed energy subspace. 
In this case the maximum of the Shannon entropy can be identified with the Boltzmann entropy 

\begin{equation}
\mathrm{max}\left( S\right) =\log W\left( E\right) , 
\end{equation}
where $W$ is the total state density with the energy E. The microcanonical case can also be seen as a particular Gibbs equilibrium (\ref{EQ:D}) for which both the energy and its fluctuation are constrained. This so called Gaussian ensemble in fact interpolates between the canonical and microcanonical ensemble depending upon the constrain on the energy fluctuation\cite{gauss,gauss2}, and 
the same procedure can be applied to any conservation law. In this sense 
the Gibbs formulation (\ref{EQ:D}) can be considered as the most general.

The ensemble of extensive variables constrained exactly or in average completely defines the 
statistical ensemble. This means that many different ensembles can be defined, 
and the most appropriate description of a finite system may be different from the standard  
 microcanonical, canonical or grand-canonical.

\subsection{Problem of the Finite Size}\label{ssec:Size}

Moreover,the  microcanonical, canonical or grand-canonical 
ensembles require the definition of boundary
condition. As discussed below, this definition is an infinite information 
incompatible with the Gibbs ideas. One can rather introduce boundaries 
through additionalconstraints $\hat{A}_{\ell }$ taking advantage of the 
general formalism above. It should be noticed that the definition of statistical ensemble forwhich the boundaries are defined using constraints such as $<\hat{r}^{2}>$ or more complex moments of the matter distribution can also be used for opensystems considered at a finite time as discussed in reference \cite{t:dep:stat,Chomaz4}.

 \begin{figure}[tbp]\resizebox{0.65\columnwidth}{!}{\ \includegraphics{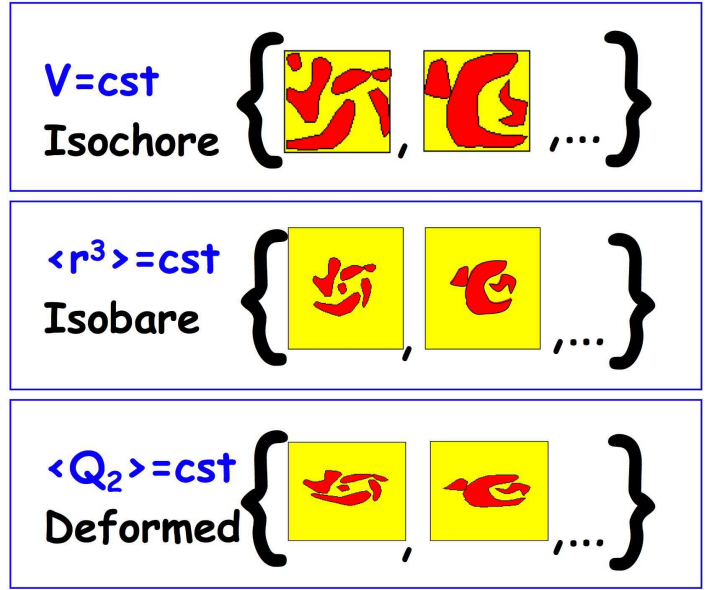} }
 \caption{Illustration of three different ensembles for a finite system: 
a system in a cubic box with a fixed volume which can be seen as an isochore ensemble; 
 a system constrained by its average size playing the role of a confining pressure, 
 which can be therefore called an isobar ensemble; and a quadrupole deformation.}
 \label{fig:3}
 \end{figure}

Figure 2 illustrates this diversity of statistical ensembles 
concerning the size observables of a finite system
 with three cases: a fixed volume with a specific shape, a size known only in average 
 and a quadrupole deformation. Many other ensembles can be defined 
 using higher multipoles as well as other observables, including 
  time odd quantities such as angular momentum $\hat{l}=\hat{r}\wedge \hat{p}$ or 
 radial flow $\hat{r}\cdot \hat{p}$. 
 Such constraints are an explicit way to describe statistical ensembles 
  associated with time evolution \cite{t:dep:stat,Chomaz4}.

\subsection{Finite size and boundary conditions}\label{sec:Finite}


The above discussion stresses the important problem when 
considering finite size systems which is the need 
to define boundary conditions to define the finite size. 
This is only a mathematical detail for ''condensed'' systems, i.e. finite size self-bound 
systems in a much larger container, or particles trapped in an external confining potential
\cite{traps}. 
In the other cases, finite size systems can only be defined when 
proper boundary conditions are specified. 
Conversely to the thermodynamic limit which, when it exists, clearly isolates bulk properties independent of the actual shape of the container, finite size systems explicitly depend on boundary conditions.

\begin{figure}[tbp]\resizebox{0.65\columnwidth}{!}
 {\ \includegraphics{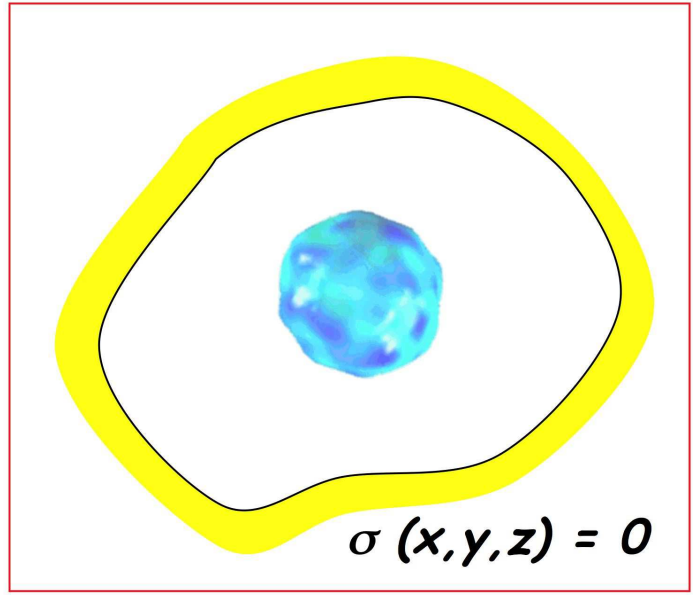} }
\caption{Schematic representation of the boundary condition around 
a nucleus computed with a mean-field approach as a 
surface defined by the implicit equation $\sigma(x,y,z)=0$ 
 which can be any shape.}\label{fig:1}
\end{figure}

From a mathematical point of view the system Hamiltonian $\hat{H}$ is not defined until boundary conditions are specified. 
For example for a particle problem a boundary can be the definition of a surface given by the implicit equation $\sigma (x,y,z)=0$.  (see figure \ref{fig:1}). 
Since the Hamiltonian $\hat{H}_{\sigma }$ 
explicitly contains the boundary, the entropy $S_{\sigma }$ also directly depends upon the definition of this boundary, according to
 
\begin{equation}
 S_{\sigma }(E)=\log  \mathrm{tr}\delta(E-\hat{H}_{\sigma }).
\end{equation}
The first conclusion is then that for a finite system the microcanonical
ensemble is ill-defined and so the thermodynamic properties of finite systems 
directly depend upon the boundary conditions, i.e. the size and shape and all small 
details of the considered container.

This brings an even more severe conceptual problem;
the knowledge of 
the boundary requires an infinite information: 
the values of the function $\sigma $ defining the actual surface in each space point. 
This is easily seen introducing the projector $\hat{P}_{\sigma}$  over the surface and its
exterior. Indeed the boundary conditions applied to each microstate 
\begin{equation}
\hat{P}_{\sigma}\left| \Psi ^{\left( n\right) }\right\rangle
=0
\end{equation}
 is exactly equivalent to the extra constraint 
\begin{equation}
<\hat{P}_{\sigma}>=\mathrm{Tr}\hat{D}\hat{P}_{\sigma}=0 \ .
\end{equation}
If we note again $\hat{\mathbf{A}}$ the observables 
characterizing a given equilibrium, the density matrix including the boundary condition constraints 
reads
 
\begin{equation}
\hat{D}_{\mathbf{\lambda }\sigma}=\frac{1}{Z_{\mathbf{\lambda }\sigma}}\exp -\mathbf{%
\ \lambda} .\hat{\mathbf{A}}-b\hat{P}_{\sigma}  \label{modif-D-2}
\end{equation}
which shows that the thermodynamics of the system 
does not only depend on the Lagrange multiplier $b$, but on the whole
surface. 
For the very same global features such as the
same average particle density or energy, we will have as many different
thermodynamics as boundary conditions.
More important, to
specify the density matrix, the projector $\hat{P}_{\sigma}$ has to be exactly
known and this is in fact impossible in actual experiments. 

Moreover, the nature of $\hat{P}_{\sigma}$ is
intrinsically different from the usual global observables $\hat{A}_{\ell }$.
At variance with the $\hat{A}_{\ell }$, $\hat{P}_{\sigma}$ is a many-body operator 
which does not correspond to any physical measurable observable.
The knowledge of $\hat{P}_{\sigma}$ requires the
exact knowledge of each point of the boundary surface while no or few
parameters are sufficient to define the $\hat{A}_{\ell }.$
If we consider statistical physics as founded by the concept of minimum 
information\cite{jaynes,balian},
it is incoherent to introduce an exact knowledge of the boundary.

One should rather apply the minimum information concept also to the 
boundaries, for example introducing a hierarchy of collective observables 
which define the size and shape of the considered system. 
This amounts to introduce statistical ensembles treating the 
boundaries as additional extensive variables fixed by conjugated Lagrange
parameters\cite{Chomaz4}. 
If for instance we consider that the relevant size information for an unbound
system is its global square radius $<\hat{R}^{2}>$,  the adequate constraint 
\begin{equation}
{-\lambda \hat{R}^2}
\end{equation}
should be introduced in the statistical treatment while all boundary conditions 
should be removed to infinity.

\section{Maximum Entropy States}

As we have discussed in the previous sections, a statistical treatment is 
best suited 
whenever a very large number of microstates exists for a given set
of observables. 
An ensemble of events coming from similary prepared initial systems 
and/or selected by sorting the final states using experimental observations 
always constitutes a statistical ensemble.
Indeed, 
if we are able to recognize all the relevant degrees of freedom 
(i.e. the observations with a strong information content) 
the ensemble of replicas is by construction 
a statistical ensemble, i.e. a Gibbs equilibrium in the extended sense 
of section \ref{sec:Stat} above. Indeed, if we are able to show that the observed 
ensemble deviates from the maximum entropy state this means that at least 
one observable give a different observation in the actual state and in the 
maximum entropy state. Then, one should introduce this observable 
as a relevant information and modify a maximum entropy state. Then, iteratively,
one is extracting the relevant observable space and at the end of the process
the actual state is then identical to the maximum entropy state. Of course nothing
is preventing that at the end of the process the whole Liouville space is needed
as a relevant information space. Then the maximum entropy state is nothing but
a complete description of any state and in this case the statistical mechanics 
tools are not useful. \footnote{A way out can be to introduce another entropy better suited to 
treat the relation between the information and the associated state as discussed in 
other contributions in this book. We will not elaborate more on this point here and rather 
refocus the discussion on the generalized Gibbs equilibria. }

However the whole idea behind the second principle of thermodynamics and
the Gibbs ansatz and the information theory is that in most physical system
the relevant information is only a small subset of the possible observations.     
%
The fundamental 
problem is then the recognition 
of  the relevant observables. A generic procedure is to make
assumptions for the relevant information and then to compare 
the maximum entropy state prediction for other observables with the
actual observations. Any important observation signals 
 a missing part of the information to be possibly introduced
as a constraint correcting the observed deviation.  


The theoretical justification of this minimal statistical picture comes from
the fact that complex classical systems subject to a non linear dynamics are 
generally mixing\cite{Cauley}.   
In such a case the statistical ensemble is created by the propagation in time of initial fluctuations. 
The averages are averages over the initial conditions and the mixing character of the dynamics 
(if it can be proved) insures that the initial fluctuations are amplified in such a way that the ensemble of events covers the whole phase space uniformly. 
\footnote{meaning that any phase space point gets close to at least one event.} 
For a classical dynamics which conserves the phase space volume of the ensemble of events, this means that the initial distribution is elongated and folded in such away that it gets close to any point of the phase space (the so-called baker transformation). This classical 
picture can be replaced in the quantum case by the idea of projection of irrelevant correlations \cite{balian2}. 
The phase space can be described as a subspace of all possible observations. 
The regular quantum dynamics in the full space is transformed into a complex dynamics
by the projection in the relevant observation sub-space. 
Then two different realizations corresponding to the same projection, i.e. the same point in 
the relevant space, may differ in the full space (and consequently in their successive 
evolution) because of the unobserved correlations. 
This ensemble of correlations may lead to a statistical ensemble of realizations 
after a finite time.
This phenomenon is often described introducing stochastic dynamics, 
i.e.assuming that the unobserved part of the dynamics which is averaged over 
is a random process
\cite{Balian3,Chomaz5}.




\subsection{Distance to equilibrium}

An important point to be discussed is the justification of the statistical description. 
As we have just mentioned,
the applicability of a statistical picture is in most cases an hypothesis (or a principle like in the thermodynamics second law). 
Therefore, the equilibrium hypothesis should be a posteriori controlled. 
Different properties can provide tests of equilibration such as
\begin{itemize}
\item  the comparison with statistical models,
\item  the consistency of thermodynamical quantities, namely the compatibility 
of the different intensive variables measurements (e.g. of the different thermometers) 
or the fulfillment of thermodynamic relations between averages and fluctuations 
(e.g. $\sigma _{A_{\ell }}^{2}=\partial^{2}\log Z_{\mathbf{\lambda }}/\partial \lambda _{\ell }^{2}=-\partial<A_{\ell }>/\partial \lambda _{\ell }$ since $<A_{\ell }>=-\partial\log Z_{\mathbf{\lambda }}/\partial \lambda _{\ell }$ )
\item  the memory loss or the independence of the results on the preparation method of the considered ensemble.
\end{itemize}
However, it should be stressed that the real question is not
whether the system is {\it at} equilibrium, but rather {\it how far} is it
from a given equilibrium?
Indeed, 
equilibrium is a theoretical abstraction which cannot be achieved in the real world
\footnote{For example, the Boltzmann law should be an exponential decay 
up to an infinite energy.}.
To answer this question we should define a distance. 
The first idea could be to use the Liouville metric 

\begin{equation}
d_{eq}^{2}=\mathrm{tr}\left( \hat{D}-\hat{D}_{\mathbf{\lambda }}\right) ^{2} 
\end{equation}
between the actual ensemble characterized by the density matrix $\hat{D}$, 
and the equilibrium one $\hat{D}_{\mathbf{\lambda }}$ 
computed for the same collective variables $<A_{\ell }>.$ 
This is a nice theoretical tool, but a rather difficult definition 
as far as experiments are concerned.

 Another possibility is to introduce 
entropy as a metric \cite{Balian3} 

\begin{equation}
d_{S}=\left| S[\hat{D}]-S[\hat{D}_{\mathbf{\lambda }}]\right| /S[\hat{D}_{\mathbf{\lambda }}] 
\end{equation}
This is a way to measure how far the system is from the maximum entropy state 
(
or in other words to measure how much information on the actual system is included in the collective variables $\{<A_{\ell }>\}$ and how much is out of the considered equilibrium looking at $(1-d_{S})$. This is a more physical distance but again it is difficult to implement in real experimental situations.

A more practical measurement of the distance to equilibrium is to focus on the information used to deduce physical properties. 
Since the information about the actual system is contained in the observations $<\hat{O}_{i}>$, 
the natural space to introduce this distance is the observation space. 
This is a formally well defined problem since considering $\mathrm{Tr}\hat{O}_{i}\hat{O}_{j}$ as the scalar product between observables, the observation space has a well defined topology. Then, when orthogonal observables are considered
\footnote{if observables are not orthogonal it is alway possible to use a Schmitt procedure to define a set of orthogonal observables\cite{balian}.},
the distance to equilibrium is simply 

\begin{equation}
d_{i}=\left| <\hat{O}_{i}>-<\hat{O}_{i}>_{eq}\right| 
\end{equation}
Of course the considered observables should not be the one used to describe the statistical equilibrium.
A typical example is given by the difference between the measured fluctuations 
$\sigma _{A_{\ell }}^{2}=<A_{\ell }^{2}>-<A_{\ell }>^{2}$
and the expected ones $\sigma _{A_{\ell }}^{2}=-\partial <A_{\ell }>/\partial \lambda _{\ell }$
in the ensemble controlled by the $\lambda _{\ell }.$

\section{Finite systems and ensemble inequivalence}

The Van Hove theorem\cite{vanhove} 
is a fundamental theorem in statistical mechanics, since it guarantees the equivalence between different statistical ensembles at the thermodynamic limit of systems presenting only finite range interactions. 
In finite systems, this equivalence can be  strongly violated in particular  
in first order phase transitions regions. 
This violation can persist up to the thermodynamic limit in the case of long range forces. 
The non-equivalence of statistical ensembles has important conceptual consequences. 
It implies that the value of thermodynamic variables for the very same system depends on the type of experiment which is performed (i.e. on the ensemble of constraints which are put on the system), 
contrary to the standard thermodynamic viewpoint that water heated in a kettle 
is the same as water put in an oven at the same temperature. 
Ensemble inequivalence is the subject of an abundant literature (see for example refs.\cite{Gulminelli,Costeniuc,Thirring,Huller,Ellis,Dauxois,Barre,Ispolatov} 
for a discussion in a general context, and refs.\cite{Kastner,Gross3,Huller2,Pleimling,Chomaz02,Chavanis}
concerning phase transitions ).

Generally speaking, for a given value of the control parameters (or \textit{intensive variables}) $\lambda_{\ell }$, the properties of a substance are univocally defined, i.e. the conjugated \textit{extensive variables} $<\hat{A}_{\ell }>$ have a unique value unambiguously defined by the corresponding equation of state 
\begin{equation}
<A_{\ell}>=-\partial_{\lambda_{\ell}}\log Z(\{\lambda_{\ell}\}
\end{equation}
In reality, this fixes only the average value and the event by event value of the observation of $\hat{A}_{\ell }$ produces a probability distribution. The intuitive expectation that extensive variables at equilibrium have a unique value therefore means that the probability distribution is narrow and normal, such that a good approximation can be obtained by replacing the distribution with its most probable value. 
The normality of probability distributions is usually assumed on the basis of the central limit theorem. 
However, in finite systems the probability distributions has a finite width and moreover it can 
depart from a normal distribution. 
We will discuss in particular the case of a bimodal distribution
\cite{topology}: in this case two different properties (phases) coexist for the same value of the intensive control variable.


The topological anomalies of probability distributions and the failure of the central limit theorem in phase coexistence imply that in a first order phase transition the different statistical ensembles are in general not equivalent and different phenomena can be observed depending on the fact that the controlled variable is extensive or intensive.
In the following we will often take as a paradigm of intensive ensembles the canonical ensemble 
for which the inverse of the temperature $\beta ^{-1}$ 
is controlled, while the archetype of the extensive ensemble will be the microcanonical one for which  energy is strictly controlled.

\subsection{Laplace versus Legendre Transformations}

The relation between the canonical entropy and the logarithm of the partition sum 
is given by a Legendre transform eq.(\ref{EQ:Legen}). 
It is important to distinguish between 
transformations within the same ensemble, as the Legendre transform, 
and transformations between different ensembles, 
which are given by non linear integral transforms\cite{Chomaz02}. 
Let us consider energy as the extensive observable and inverse temperature 
$\beta $ as the conjugated intensive one. The definition of the canonical partition sum is 

\begin{equation}
Z_{\beta }=\sum_{n}\exp (-\beta E^{(n)}) ,
\label{partition_sum}
\end{equation}
where the sum runs over the available eigenstates $n$ of the Hamiltonian. Here, we assume that the partition sum converges; this is not always the case as discussed in ref.\cite{ang_mom}.
The possible divergence of the thermodynamic potential of the intensive ensemble 
is already a known case of ensemble inequivalence\cite{gross-here,ang_mom}. 
Computing the canonical (Shannon) entropy we get 

\begin{equation}
S_{can}(<E>)=\log Z_{\beta }+\beta <E>  ,
\label{EQ:Legendre-can}
\end{equation}
which is an exact Legendre transform since the EoS reads $<E>=-\partial_{\beta }\log Z_{\beta }.$ 
If energy can be treated as a continuous variable, eq.(\ref{partition_sum}) can be written as:

\begin{equation}
Z_{\beta }=\int_{0}^{\infty }dE\ W(E)\exp (-\beta E),  
\label{laplace}
\end{equation}
where energies are evaluated from the ground state. 
Eq.(\ref{laplace}) is a Laplace transform between the canonical partition sum and 
the microcanonical density of states linked to the entropy by $S_{E}=\ln W(E)$. 
If the integrand $f(E)$ $=$ $\exp (S_E-\beta E)$ is a strongly peaked function, 
it can be approximated by a gaussian (saddle point approximation) 
so that the integral can be replaced by the maximum $f(\bar{E})$ times a Gaussian integral. 
Neglecting this factor we get 

\begin{equation}
Z_{\beta }\approx W(\bar{E})\exp (-\beta \bar{E})  ,
\label{saddle}
\end{equation}
which can be rewritten as 

\begin{equation}
\ln Z_{\beta }\approx S_{\bar{E}}-\beta \bar{E};  
\label{legendre_approx}
\end{equation}
or introducing the free energy $F_{T}=-\beta ^{-1}\ln Z_{\beta }$ 

\begin{equation}
F_{T}\;\approx \;\bar{E}-TS_{\bar{E}} .
\end{equation}
Eq.(\ref{legendre_approx}) has the structure of an approximate Legendre transform similar to the exact 
espression (\ref{EQ:Legendre-can}).
This shows that in the lowest order saddle point approximation eq.(\ref{saddle}), 
the ensembles differing at the level of constraints acting on a 
specific observable (here energy) lead to the same entropy, i.e. they are equivalent. 
We will see in the next section that however the saddle point approximation eq.( \ref{saddle}) can be highly incorrect close to a phase transition for the simple reason that the integrand is bimodal making a unique saddle point approximation inadequate. In this case eq.(\ref{legendre_approx}) cannot be applied, eq.(\ref{laplace}) is the only correct transformation between the different ensembles, and ensemble inequivalence naturally arises.

To summarize one should not confuse
\begin{itemize}
\item  the link between the thermodynamical potential of the intensive (e.g.log of canonical partition sum) and of the extensive ensemble (e.g. themicrocanonical entropy) which are always related with a Laplace transform.This Laplace transform may lead to an approximate Legendre transformationfor normal distributions but we know that this Legendre transformation isonly approximate and might be present strong deviations if the distributionis abnormal.
\item  with the exact Legendre transform between the entropy of theintensive ensemble and the corresponding thermodynamical potential.\end{itemize}This simply corresponds to the fact that the microcanonical and canonicalentropies can be very different.

\subsection{ Ensemble Inequivalence in Phase Transition Regions}

Let us consider the case of a first order phase transition where the 
canonical energy distribution 

\begin{equation}
P_{\beta _{0}}\left( E\right) =W(E)\exp (-\beta _{0}E)/Z_{\beta _{0}}
\label{EQ:PBoltzman}
\end{equation}
has a characteristic bimodal shape \cite{topology,binder,labastie} 
at the temperature $\beta _{0}$ with two maxima $\overline{E}_{\beta }^{\left(1\right) }$, $\overline{E}_{\beta }^{\left( 2\right) }$ that can be associated with the two phases.
It is easy to see that eq.(\ref{laplace}) 
can also be seen as a Laplace transform of the canonical probability 
$P_{\beta _{0}}\left( E\right) $ 

\begin{equation}
Z_{\beta }=Z_{\beta _{0}}\int_{0}^{\infty }dE\ P_{\beta _{0}}\left( E\right)\exp (-(\beta -\beta _{0})E).
\end{equation}
A single saddle point approximation is not valid when $P_{\beta _{0}}\left(E\right) $ is bimodal; 
however it is always possible to write 

\begin{equation}
P_{\beta }=m_{\beta }^{\left( 1\right) }P_{\beta }^{\left( 1\right) }+m_{\beta}^{\left( 2\right) }P_{\beta }^{\left( 2\right) },
\end{equation}
with $P_{\beta }^{\left(i\right) }$ mono-modal normalized probability distribution peaked at $\overline{E}_{\beta }^{\left( i\right) }$. 
The canonical mean energy is then the weighted average of the two energies

\begin{equation}
\left\langle E\right\rangle _{\beta }=\tilde{m}_{\beta }^{\left( 1\right) }\overline{E}_{\beta }^{\left( 1\right) }+\tilde{m}_{\beta }^{\left(2\right)}\overline{E}_{\beta }^{\left( 2\right) } , 
\label{EQ:_E_can_backbend}
\end{equation}
with 
\begin{equation}
\tilde{m}_{\beta }^{\left( i\right) }=m_{\beta }^{\left( i\right)}\int dEP_{\beta }^{\left( i\right) }(E)E/\overline{E}_{\beta }^{\left(i\right) }\simeq m_{\beta }^{\left( i\right) }.
\end{equation}
Since only one mean energy is associated with a given temperature $\beta ^{-1}$, 
the canonical caloric curve is monotonous, and the microcanonical one is not. 
Indeed it is immediate to see from eq.(\ref{EQ:PBoltzman}) that the 
bimodality of $P_{\beta }$ implies then a back bending of the microcanonical 
caloric curve $T^{-1}=\partial _{E}S$, meaning that in the first order phase 
transition region the two ensembles are not equivalent.
If instead of looking at the average $\left\langle E\right\rangle _{\beta }$
we look at the most probable energy $\overline{E}_{\beta }$ , 
this (unusual)canonical caloric curve is identical to the microcanonical one, 
up to the transition temperature $\beta _{t}^{-1}$ for which the two components 
of $P_{\beta }\left( E\right) $ have the same height. 
At this point the most probable energy jumps from the low to the high energy 
branch of the microcanonical caloric curve. 

The question arises whether this violation of ensemble equivalence survives 
towards the thermodynamic limit. 
This limit can be expressed as the fact 
that the thermodynamic potentials per particle converge 
when the number of particles $N$ goes to infinity : 

\begin{equation}
f_{N,\beta }=\beta ^{-1}\frac{\log Z_{\beta }}{N}\rightarrow \bar{f}_{\beta } \; ; \;  
s_{N}\left( e\right) =\frac{S(E)}{N} \rightarrow \bar{s}\left( e\right) 
\end{equation} 
where $e=E/N$. 
Let us also introduce the reduced probability $p_{N,\beta }\left( e\right) =\left( P_{\beta }(N,E)\right)^{1/N}$ which then converges towards an asymptotic distribution 

\begin{equation}
p_{N,\beta}\left( e\right) \rightarrow \bar{p}_{\beta }\left(e\right) \; ;\; 
\bar{p}_{\beta }\left( e\right) =\exp \left( \bar{s}(e)-\beta e+\bar{f}_{\beta}\right).
\end{equation}
Since $P_{\beta }(N,E)\approx $ $\left( \bar{p}_{\beta }\left(e\right) \right) ^{N}$, 
one can see that when $\bar{p}_{\beta }\left(e\right) $ is normal, 
the relative energy fluctuation in $P_{\beta }(N,E)$ is suppressed by a factor $1/\sqrt{N} $. 
At the thermodynamic limit $P_{\beta}$ reduces to a $\delta $-function 
and ensemble equivalence is recovered. 
To analyze the thermodynamic limit of a  bimodal $p_{N,\beta }\left( e\right) $, 
let us introduce as before $\beta _{N,t}^{-1}$ the temperature for which the two 
maxima of $p_{N,\beta }\left( e\right) $ have the same height. 
For a first order phase transition $\beta _{N,t}^{-1}$ converges to a fixed point 
$\bar{ \beta}_{t}^{-1}$ as well as the two maximum energies 
$e_{N,\beta }^{\left(i\right) } \rightarrow \bar{e}_{\beta}^{\left( i\right) }$. 
For all temperatures lower (higher) than $\bar{\beta} _{t}^{-1}$ 
only the low (high) energy peak will survive at the thermodynamic limit, 
since the difference of the two maximum probabilities will be raised to the power $N.$ 
Therefore, below $\bar{e}_{\beta }^{\left( 1\right) }$ and above 
$\bar{e}_{\beta}^{\left( 2\right) }$ the canonical caloric curve coincides 
with the microcanonical one in the thermodynamic limit. 
In the canonical ensemble the temperature $\bar{\beta}_{t}^{-1}$ corresponds to a 
discontinuity in the state energy irrespectively of the behavior of the entropy 
between $\bar{e}_{\beta }^{\left( 1\right) }$ and $\bar{e} _{\beta }^{\left( 2\right) }$.

The microcanonical caloric curve in the phase transition region may either
converge towards the Maxwell construction, or keep a backbending behavior\cite{leyvraz}, 
since a negative heat capacity system can be thermodynamically stable even 
in the thermodynamic limit if it is isolated \cite{Thirring}. 
Examples of a backbending behavior at the thermodynamic limit 
have been reported for a model many-body interaction taken as a functional 
of the hypergeometric radius in the analytical work of ref.\cite{lynden-bell}, 
and for the long range Ising model \cite{ruffo}. 
This can be understood as a general effect of long range interactions 
for which the topological anomaly leading to the convex intruder in the entropy 
is not cured by increasing the number of particles\cite{ruffo,pettini}. 
Conversely, for short range interactions \cite{gross} 
the backbending is a surface effect which should disappear at the thermodynamic limit. 
This is the case for the Potts model\cite{Gross3}, the microcanonical model of 
fragmentation of atomic clusters\cite{gross-clus} and for the lattice gas 
model with fluctuating volume\cite{europhys} . 
The interphase surface entropy goes to zero as $N\rightarrow \infty $ 
in these models, leading to a linear increase of the entropy in agreement with the canonical predictions.
Within the approach based on the topology of the probability distribution of observables \cite{topology} it was shown that ensemble inequivalence arises from fluctuations of the order parameter \cite{Gulminelli}. Ensembles putting different constraints on the fluctuations of the order parameter lead to a different thermodynamics. In the case of phase transitions with a finite latent heat, the total energy usually plays the role of an order parameter except in the microcanonical ensemble which therefore is expected to present a different thermodynamics than the other ensembles\cite{gross-here}. This inequivalence may remain at the thermodynamic limit if the involved phenomena are not reduced to short range effects. 

\subsection{Temperature jump at constant energy}

\begin{figure}[tbp]\resizebox{0.75\columnwidth}{!}{\ 
\includegraphics{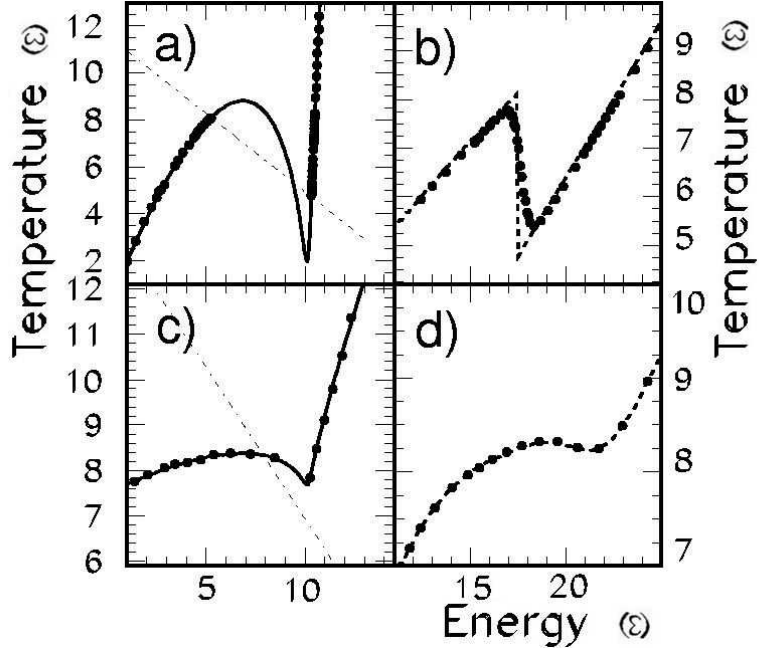} }
\caption{Left panels: temperature as a function of the potential energy $E_{2}$ (full lines) and of the kinetic energy $E-E_{2}$ (dot-dashed lines)for two model equation of states of classical systems showing a first order phase transition. Symbols: temperatures extracted from the most probable kinetic energy thermometer from eq.(\ref{thermometer}). 
Right panels: total caloric curves (symbols) corresponding to the left panels and thermodynamic 
limit of eq.(\ref{zero_princ}) (dashed lines).}
\label{fig:ineq-1}
\end{figure}

In particular, it may happen that the energy of a subsystem becomes an order parameter when the total energy is constrained by a conservation law or a microcanonical sorting.
This frequently occurs for Hamiltonians containing a kinetic energy contribution\cite{lynden-bell,ruffo,Dauxois2}: 
if the kinetic heat capacity is large enough, it becomes an order parameter in the microcanonical ensemble. 
Then, the microcanonical caloric curve presents at the thermodynamical limit a temperature jump in complete disagreement with the canonical ensemble.

To understand this phenomenon, let us consider a finite system for which the Hamiltonian can be separated into two components $E=E_{1}+E_{2},$ that are statistically independent ($W(E_{1},E_{2})=W_{1}(E_{1})W_{2}(E_{2})$) and such that the associated 
degrees of freedom scale in the same way with the number of particles; 
we will also consider the case where $S_{1}=\log W_{1}$ has no anomaly while $S_{2}=\log W_{2}$ presents a convex intruder\cite{gross} which is preserved at the thermodynamic limit. 
Typical examples of $E_{1}$ are given by the kinetic energy for a classical system with velocity independent interactions, or other similar one body operators \cite{ruffo}.
The probability to get a partial energy $E_{1}$ when the total energy is $E$ is given by

\begin{equation}
P_{E}\left( E_{1}\right) =\exp \left( S_{1}\left( E_{1}\right) +S_{2}\left(E-E_{1}\right) -S\left( E\right) \right)  \label{eq:P1}
\end{equation}

The extremum of $P_{E}\left( E_{1}\right) $ is obtained for the partitioning of the total energy $E$ between the kinetic and potential components that equalizes the two partial temperatures

\begin{equation}
 \overline{T_{1}}^{-1}=\partial_{E_{1}}S_{1}( \overline{E}_{1})=\partial _{E_{2}}S_{2}(E-\overline{E}_{1})= \overline{T_{2}}^{-1}.
\label{thermometer}
\end{equation}
 
If $\overline{E}_{1}$is unique, $P_{E}\left(E_{1}\right) $ is mono-modal and we can use a saddle point approximation around this solution to compute the entropy 

\begin{equation}
S\left( E\right) =\log\int_{-\infty }^{E}dE_{1}\exp \left( S_{1}\left( E_{1}\right) +S_{2}\left(E-E_{1}\right) \right) .
\end{equation}

 At the lowest order, the entropy is simply additive so that the microcanonical temperature of the global system $\partial _{E}S(E)=\overline{T}^{-1}$ is the one of the most probable energy partition. Therefore, the most probable partial energy $\overline{E}_{1}$ acts as a microcanonical thermometer. If $\overline{E}_{1}$ is always unique, the kinetic thermometer in the backbending region will follow the whole decrease of temperature as the total energy increases. Therefore, the total caloric curve will present the same anomaly as the potential one. If conversely the partial energy distribution is double humped \cite{berry},
 then the equality of the partial temperatures admits three solutions, one of them $\overline{E}_{1}^{\left( 0\right) }$ being a minimum. At this point the partial heat capacities
 
 \begin{equation}
  C_{1}^{-1}=-\overline{T}^{2}\,\partial_{E_{1}}^{2}S_{1}(\overline{E}_{1}^{(0)})\; ;\; C_{2}^{-1}=-\overline{T}^{2}\,\partial _{E_{2}}^{2}S_{2}(E-\overline{E}_{1}^{(0)})
  \end{equation}
   
  fulfill the relation 

\begin{equation}
C_{1}^{-1}+C_{2}^{-1}<0  \label{eqcentral}
\end{equation}

This happens when the potential heat capacity is negative and the kinetic energy is large enough ($C_{1}>-C_{2}$) to act as an approximate heat bath: the partial energy distribution $P_{E}\left( E_{1}\right) $ in the microcanonical ensemble is then bimodal as the total 
energy distribution $P_{\beta }\left( E\right) $ in the canonical ensemble, implying that the kinetic energy is the order parameter of the transition in the microcanonical ensemble. In this case the microcanonical temperature is given by a weighted average of the two estimations from the two 
maxima of the kinetic energy distribution 

\begin{equation}
T=\partial _{E}S(E)=\frac{\overline{P}^{(1)}\sigma ^{(1)}/\overline{T}^{\left( 1\right) }+\overline{P}^{(2)}\sigma ^{(2)}/\overline{T}^{\left(2\right) }}{\overline{P}^{(1)}\sigma ^{(1)}+\overline{P}^{(2)}\sigma ^{(2)}}
\label{zero_princ}
\end{equation}

where $\overline{T}^{\left( i\right) }=T_{1}(\overline{E}_{1}^{\left(i\right) })$ are the kinetic temperatures calculated at the two maxima, $\overline{P}^{\left( i\right) }=P_{E}(\overline{E}_{1}^{\left( i\right) })$ are the probabilities of the two peaks and $\sigma ^{\left( i\right) }$ their widths. 
At the thermodynamic limit eq.(\ref{eqcentral}) reads $c_{1}^{-1}+c_{2}^{-1}<0$, with $c=\lim_{N \to \infty }C/N$ . If this condition is fulfilled, the probability distribution $P_{\beta}(E)$ presents two maxima for all finite sizes and only the highest peak survives at $N=\infty $ . Let $E_{t}$ be the energy at which $P_{E_{t}}(\overline{E}^{\left( 1\right) })=P_{E_{t}}(\overline{E}^{\left( 2\right) })$. Because of eq.(\ref{zero_princ}), at the thermodynamic limit the caloric curve will follow the high (low) energy maximum of $P_{E}\left( E_{1}\right) $ for all energies below (above) $E_{t}$; there will be a temperature jump at the transition energy $E_{t}.$ Let us illustrate the above results with two examples for a classical gas of interacting particles. For the kinetic energy contribution we have $S_{1}(E)=c_{1}\ln (E/N)^{N}$ with a constant kinetic heat capacity perparticle $c_{1}=3/2$. For the potential part we will take two polynomial parametrization of the interaction caloric curve presenting a back bending which are displayed in the left part of figure \ref{fig:ineq-1} in units of an arbitrary scale $\epsilon $. If the decrease of the partial temperature $T_{2}(E_{2})$ is steeper than $-2/3$ (figure \ref{fig:ineq-1}a ) \cite{lynden-bell} eq.(\ref{eqcentral}) is verified and the kinetic caloric curve $T_{1}(E-E_{1})$(dot-dashed line) crosses the potential one $T_{2}(E_{2})$ (full line) in three different points for all values of the total energy lying inside the region of coexistence of two kinetic energy maxima. The resulting caloric curve for the whole system is shown in figure \ref{fig:ineq-1}b (symbols) 
together with the thermodynamic limit (lines) evaluated from the double saddle point approximation (\ref{fig:ineq-1}). In this case one observes a temperature jump at the transition energy. If the temperature decrease is smoother (figure \ref{fig:ineq-1}c) the shape of the interaction caloric curve is preserved at the thermodynamic limit (figure \ref{fig:ineq-1}d).

\begin{figure}[tbp]\resizebox{0.75\columnwidth}{!}{\ \includegraphics{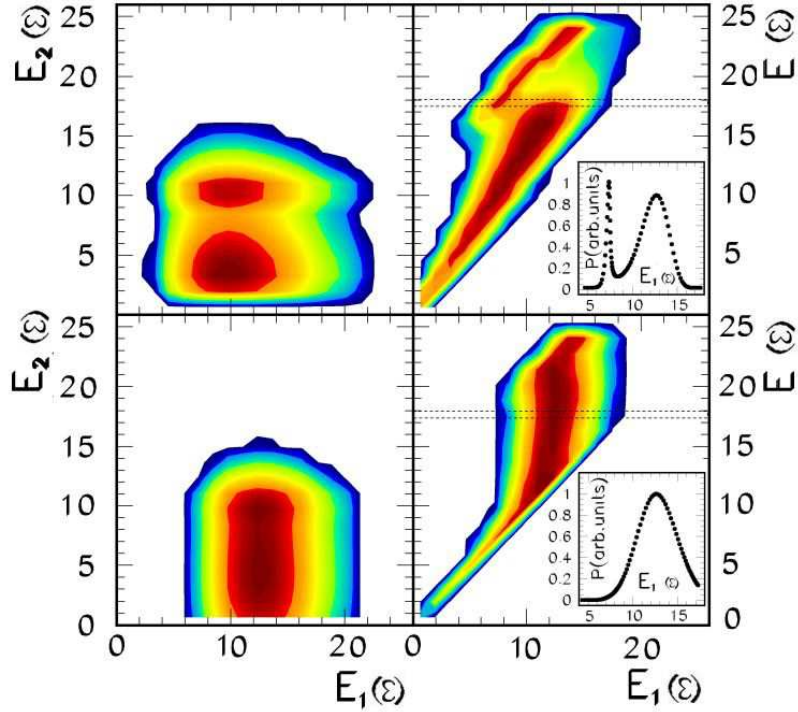} }
\caption{Canonical event distributions in the potential versus kinetic 
energy plane (left panels) and total versus kinetic energy plane (right panels) at the transition temperature for the two model equations of state of figure \ref{fig:ineq-1}. The inserts show two constant total energy cuts of the distributions.}
\label{fig:ineq-2}
\end{figure}

The occurrence of a temperature jump in the thermodynamic limit is easily spotted by looking at the bidimensional canonical event distribution $P_{\beta }(E_{1},E_{2})$
shown at the transition temperature $\beta =\beta _{t}$  in the left part of figure \ref{fig:ineq-2} for the two model equation of states of figure \ref{fig:ineq-1}. 
In the canonical ensemble 
the kinetic energy distribution is normal. 
These same distributions are shown
as a function of $E$ and $E_{1}$, $P_{\beta }(E,E_{1})\propto \exp S_{1}(E_{1})\exp S_{2}(E-E_{1})\exp (-\beta E)$ in the right part of figure \ref{fig:ineq-2}. 
The microcanonical ensemble is a constant energy cut of $P_{\beta }(E,E_{1})$, which leads to the microcanonical distribution $P_{E}(E_{1})$ within a normalization constant. 
If the anomaly in the potential equation of state is sufficiently important, the distortion of 
events due to the coordinate change is such that one can still see the two phases coexist even after a sorting in energy. 

\section{\protect\smallskip Definitions of Phase Transitions in Finite Systems}

Phase transitions are universal properties of matter in interaction. In macroscopic physics, they are singularities (i.e. non-analitical behaviors) in the system equation of state (EoS) and hence classified according to the degree of non-analyticity of the EoS at the transition point. Then, a phase transition is an intrinsic property of the system and not of the statistical ensemble used to describe the equilibrium. Indeed, at the thermodynamic limit all the possible statistical ensembles converge 
towards the same EoS 
and the various thermodynamic potentials are related by simple Legendre transformations leading to a unique thermodynamics.
On the other side for finite systems, as discussed above, two ensembles which put different constraints on the fluctuations of the order parameter lead to qualitatively different EoS close to a first order phase transition \cite{gauss,gross}. Thermodynamic observables like heat capacities can therefore be completely
different depending on the experimental conditions of the measurement.
Moreover, such inequivalences may survive at the thermodynamic limit if forces are long ranged 
as for self gravitating objects\cite{lynden-bell,ruffo}. In fact the characteristic of phase transitions in finite systems, and in particular the occurrence of a negative heat capacity, have first been discussed in the astrophysical context\cite{Lynden-Bell2,Ispolatov,Thirring,Antonov,Hertel,Chavanis2,Padhmanaban,Katz}. 
Since
these pioneering works in astrophysics, an abundant literature is focused on the understanding of phase transition in small systems from a general point of view \cite{gross,Barre,Chomaz02,Promberger,Behringer,Kastner3,Franzosi2,Franzosi,Naudts} or in the mean-field context \cite{ruffo,Antoni} or for some specific systems such as metallic clusters \cite{labastie,berry} 
or nuclei \cite{Gulminelli2} and even DNA \cite{Wynveen}.

\subsection{Phase transitions in infinite systems}

\begin{figure}[tbp]
\resizebox{.5\columnwidth}{!}{\ \includegraphics{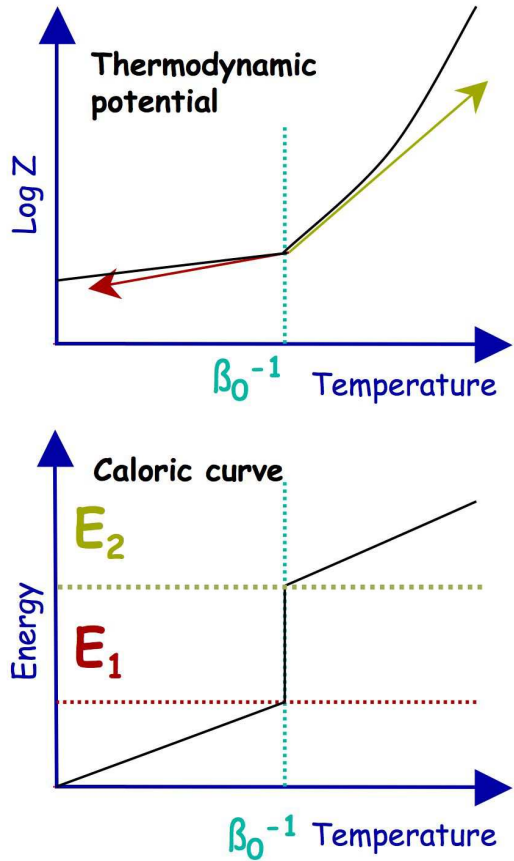} }
\caption{Schematic representation of a first order phase transition in the canonical case. 
Top: the log of the canonical partition sum (i.e. the free energy) presents an angular point 
Bottom: the first derivative as a function of the temperature 
(i.e. the energy) presents a jump. }\label{fig_phase_trans}
\end{figure}

Let us first recall the definition of phase transitions in infinite systems.
At the thermodynamic limit for short range interactions the statistical 
ensembles are equivalent and it is enough to reduce the discussion to the ensemble where 
only one extensive variable $A_L$ is kept fixed, all the others being constrained
through the associated Lagrange parameters. 
The typical example is the grandcanonical 
ensemble where only the volume $A_{L}=V$ is kept as an extensive variable.
Then all the thermodynamics is contained in the 
associated potential $\log $ $Z_{\lambda _{1,}...,\lambda _{L-1}}(A_{L})$. 
Since it is extensive, the potential is proportional to the 
remaining extensive variable 

\begin{equation}
\log Z_{\lambda_{1,}...,\lambda _{L-1}}(A_{L})=
A_{L}\lambda _{L}(\lambda _{1,}...,\lambda _{L-1})
\end{equation} 
so that all the non trivial thermodynamic 
properties are included in the reduced potential i.e. the intensive variable 

\begin{equation}
\lambda _{L}=\partial _{A_{L}}\log Z_{\lambda _{1,}...,\lambda _{L-1}}(A_{L})
=\frac{\log Z_{\lambda _{1,}...,\lambda _{L-1}}}{A_{L}}
\end{equation}
associated with $A_L$. 
In the grand canonical case $A_{L}=V$, 
the reduced potential is the pressure, $\lambda _{L}\propto P$, 
which is then a function of the temperature and the chemical potential(s).
In this limit all the thermodynamics is included in the single function 
$\lambda _{L}(\lambda _{1,}...,\lambda _{L-1})$, and this is why in the literature 
$p(V)$ is often loosely refered to as "`the"' EoS, and the existence 
of many EoS is ignored.   
If this EoS is analytical, 
all the thermodynamic quantities which are all derivatives of the thermodynamic potential,
present smooth behaviors, and no phase transition appears. 
A phase transition is a major modification of the macrostate properties for a small 
modification of the control parameters $(\lambda _{1,}...,$ $\lambda _{L-1})$%
. Such an anomalous behavior can only happen if the thermodynamic potential presents a singularity. 
This singularity can be classified according to the order of the derivative which presents a discontinuity or a divergence. According to Ehrenfest this is the order of the phase transition. 
In modern statistical mechanics, 
all the higher order are called under the generic name of continuous transitions. 
Figure \ref{fig_phase_trans} schematically illustrates a first order phase transition in the canonical ensemble.

\subsection{\protect\smallskip Phase transitions in finite systems}

\begin{figure}[tbp]\resizebox{0.7\columnwidth}{!}{\ 
\includegraphics{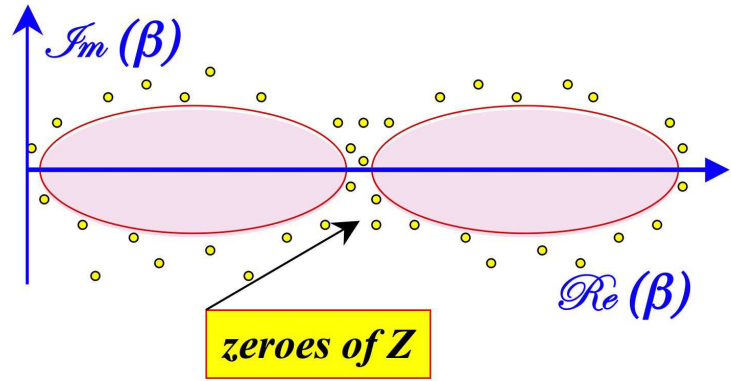} }
\caption{Schematic representation of the zeroes of the partition sum $Z$ in the complexe temperature plane. The regions where no zeroes are coming close to the real axis when the thermodynamic limit is taken will not present singularities of $\log Z.$}
\label{Fig:zeroes}
\end{figure}

As soon as one considers a finite physical system, all the above discussion does not apply. 
First the thermodynamic potential and observables are not additive, therefore 
we cannot introduce a reduced potential
and we get 

\begin{eqnarray}{\lambda}_{L}(\lambda _{1,}...,\lambda _{L-1,}A_{L}) 
&\equiv &\frac{\partial \log Z_{\lambda _{1,}...,\lambda _{L-1}}(A_{L})}{\partial A_{L}} 
\nonumber \\
&\neq &\frac{\log Z_{\lambda _{1,}...,\lambda _{L-1}}(A_{L})}{A_{L}}
\end{eqnarray}
%
i.e. the grand potential per unit volume does not give the 
pressure any more, and presents a non-trivial volume dependence. 
Moreover the analysis of the singularities of the thermodynamic potential 
has no meaning, since it is an analytical function. 
The standard statistical physics textbooks thus conclude that rigourously speaking there is no 
phase transitions in finite systems. 
However, as we have already mentioned, first for self-gravitating objects \cite{Lynden-Bell2,Thirring,Ispolatov,Antonov,Hertel,Chavanis2,Padhmanaban,Katz} 
and then in small systems \cite{lynden-bell,gross,Barre,labastie,Chomaz02,Promberger,Behringer,Kastner3,Naudts,Gulminelli2} 
it was shown that phase transitions might be associated with the occurrence of negative microcanonical heat capacities.
This can be generalized to the occurence of an inverted curvature of the thermodynamic 
potential of any ensemble keeping at least one extensive variable $A_{L}$ not orthogonal to the order parameter 
\footnote{orthogonality is here defined using the trace as a scalar product between observables
following section \ref{sec:Stat}}
 \cite{Chomaz02,Gulminelli3}. 
 In the following we call this ensemble an extensive ensemble. Then, negative compressibility or negative susceptibility should be, like negative heat capacity, observed in first order phase transitions of finite systems. In the microcanonical ensemble of classical particles, it was proposed that anomalously large fluctuations of the kinetic energy, i.e. larger then the the expected canonical value, highlight a negative heat capacity\cite{Chomaz}.
 It was then demonstrated that those two signals of a phase transition, negative curvatures and anomalous fluctuations, observed in extensive ensemble where the order parameter is fixed, are directly related to the appearance of bimodalities in the distribution of this order parameter in the intensive ensemble where the order parameter is only fixed in average through its conjugated Lagrange multiplier\cite{topology,Schmidt}. 
 The occurrence of bimodalities is discussed in the literature since a long time 
 and is often used as a practical way to look for phase transition in numerical simulations \cite{Hill,binder}; however, the general equivalence between negative curvatures and bimodalities was presented in ref.\cite{topology}. For intensive ensembles, since the pioneering work of Yang and 
 Lee\cite{Yang} another definition was proposed considering the zeroes of the partition sum in 
 the complex intensive parameter plane \cite{Yang,grossmann}. 
 The idea is simple: the zeroes of $Z$ are the singularities of $\log Z$ and so phase 
 transitions, which are singularities, must come from the zeroes of the partition sum. 
 In a finite system the zeroes of the partition sum cannot be on the real axis since the partition sum $Z$ is the sum of exponential factors which cannot produce a singularity of $\log Z.$ However, the thermodynamic limit of an infinite volume may bring the singularity on the real axis. This is 
 schematically illustrated in fig. \ref{Fig:zeroes}. 
 Only regions where zeroes converge towards the real axis may present phase transitions, while the other regions present no anomalies. The order of the transition can be associated to the asymptotic
 behavior of zeroes\cite{grossmann}.

The distribution of zeroes 
has been analyzed in ref.\cite{Lee} where the transition was studied with a parabolic entropy. In ref. \cite{Chomaz3} the equivalence of the expected behavior of the zeroes in a first order phase transition case and the occurrence of bimodalities in the distribution of the associated extensive parameter was demonstrated. 
To be precise, in this demonstration bimodality means that
the extensive variable distribution can be splitted at the transition point into 
two distributions of equal 
height, with the distance between the two maxima scaling like the system size\cite{touchette-comment}. 

\begin{figure}[tbp]
\resizebox{.5\columnwidth}{!}
{\ \includegraphics{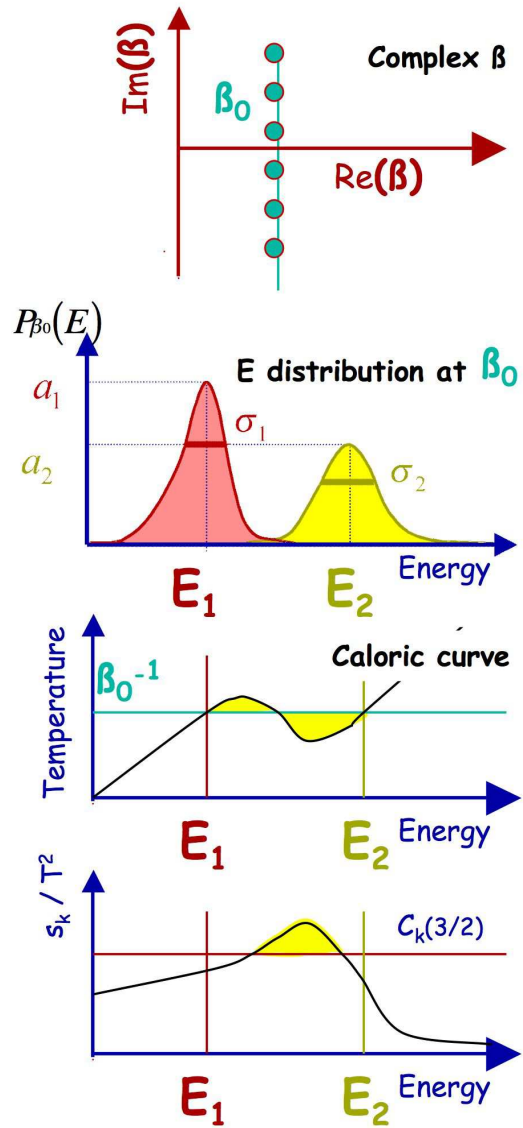} }
\caption{Schematic representation of the different equivalent definitions of first order 
phase transitions in finite systems. From top to bottom: 
the partition sum's zeroes aligning perpendicular to the real temperature axis with a density scaling like the number of particles; the bimodality of the energy distribution with a distance between the two maxima scaling like the number of particles times the latent heat; the appearence of a back-bending in the microcanonical caloric curve i.e. a negative heat capacity region; and the observation of anomalously large fluctuations of the energy splitting between the kinetic part and the interaction part. }
\label{Fig:equiv}
\end{figure}

This global picture of phase transitions in finite systems 
is summarized in figure \ref{Fig:equiv} in the case where energy is the order parameter 
of the transition. 
The occurrence of a bimodal distribution of the extensive parameter (e.g. energy) in the associated intensive (e.g. canonical) ensemble is a necessary and sufficient condition to 
asymptotically get the distribution of the Yang-Lee zeroes in the complex lagrange multiplier (e.g. temperature) plane, which is 
expected in a first order transition. 
The direction of bimodality is the direction of the order parameter.  This bimodality 
is also equivalent to the presence of an anomalous curvature in the thermodynamic potential of the extensive (microcanonical) ensemble %
obtained constraining the bimodal observable to a fixed value. 
In the extensive ensemble, the inverted curvature can be spotted looking 
for anomalously large fluctuations  (e.g. larger then the canonical ones) 
of the partition of the extensive variable (e.g. energy)  between two independent subsystems.

\section{Statistical description of evolving systems}\smallskip 

A major issue in the statistical treatment of finite systems is that most of the time open and transient systems are studied. Therefore, they are not only finite in size but also finite in time, and in fact they are evolving. 
The number of degrees of freedom of a quantum many-body problem being infinite,
it is impossible to have all the information needed to solve exactly the dynamical problem.
Since only a small part of the observation space is relevant, this time evolution may also be treated
with statistical tools. This is the purpose of many models: from Langevin approaches to Fokker-Planck equations, from hydrodynamics to stochastic Time Dependent Hartree Fock theory. 
The purpose of this chapter is not to review those  theoretical approaches, therefore we will not enter here into details about the different recent progresses, and we will rather focus this discussion on general arguments of time dependent statistical ensembles \cite{balian2,Chomaz4}. 

A statistical treatment of a dynamical process is based on the idea 
that at any time one can consider only the relevant variables $A_{\ell }$, disregarding all the other ones $a_{m}$ as irrelevant. If only the maximum entropy state is followed in time assuming that all the irrelevant degrees of freedom have relaxed instantaneously, one gets a generalized mean-field approach \cite{Balian3}. If the fluctuations of the irrelevant degrees of freedom are included, this leads to a Langevin dynamics%
\cite{Chomaz5}. With those considerations one can see that statistical approaches can be always improved including more and more degrees of freedom to asymptotically become exact. 
However, before including a huge number of degrees of freedom one should ask himself if only a few observables can take care of the most important dynamical aspects of the systems we are looking at. In a recent paper\cite{Chomaz4} it was proposed to introduce observations at different times (e.g. different freeze-out/equilibration times) as well as time odd extensive parameters. The idea is simple : maximizing the Shannon entropy with different observables $\hat{A}_{\ell }$ known at different times $t_{\ell }=t_{0}+\Delta t_{\ell }$ is a way to treat a part of the dynamics. Going to the Heisenberg 
representation, if we propagate all the $\hat{A}_{\ell }$ to the same time $t_{0}$ we get:  

\begin{eqnarray}
\hat{A}_{\ell }(t_{0}) &=&e^{-i\Delta t_{\ell }\hat{H}}\hat{A}_{\ell }e^{i%
\Delta t_{\ell }\hat{H}} \\&=&\hat{A}_{\ell }-i\Delta t_{\ell }[\hat{H},\hat{A}_{\ell }]+...
\end{eqnarray}

This shows that the time propagation introduces new constraining operators 

\begin{equation}
\hat{B}_{\ell }=-i[\hat{H},\hat{A}_{\ell }].  \label{EQ:B}
\end{equation}

If $\hat{A}_{\ell }$ is a time even observable,  $\hat{B}_{\ell }$ is a time odd operator.
Let us take the example of  an unconfined finite system characterized at a given time by a typical size $<\hat{R}^{2}>=<\hat S>,$ where $\hat{R}^{2}$ is the one body operator $\sum_{i}$ $\hat{r}_{i}^{2}$. If the whole information is assumed to be known at the same time, then the statistical distribution of event reads in a classical canonical picture 

\begin{equation}
p^{\left( n\right) }=\frac{1}{Z}e^{-\beta E^{(n)}-\lambda S^{\left( n\right)}}
\end{equation}

which is formally equivalent to a particle in a harmonic potential. 
However, if now we assume that the size information is coming from a different time, then 
according to Eq. (\ref{EQ:B}) we must introduce a new time odd operator $\hat{v}_{r}=-i[\hat{H},\hat{r}^{2}]$. For a local interaction, this reduces to 

\begin{equation}
\hat{v}_{r}=(\hat{r}\hat{p}+\hat{p}\hat{r})/m
\end{equation}

which represents a radial flow. Then the classical canonical probability reads

\begin{equation}
p^{\left( n \right) }=\frac{1}{Z}e^{-\beta \left( \mathbf{p}^{(n)}
-h(t) \mathbf{r^{(n)}} \right)^{2}-\lambda S^{\left( n\right) }}
\label{flow}
\end{equation}

which is a statistical ensemble of particles under a Hubblian flow. 
In the ideal gas model eq.(\ref{flow}) provides the exact solution at any time of the dynamics. 
This simple example shows that information theory allows treat in a statisticl picture
dynamical processes where observables are defined at different times, by
taking into account time odd components such as flows. 
This might be a tool to extract thermodynamical quantities from complex dynamics. 
In particular, the above example shows that in an open system an initial extension in space 
is always transformed into an expansion, meaning that flow is an essential ingredient 
even in statistical approaches.

\smallskip
\section{Conclusion}
\smallskip 

In conclusion, we have presented in this paper the actual understanding of the thermodynamics of finite systems from the point of view of information theory. We have put some emphasis on first order phase transitions which are
associated to specific and intriguing phenomena as bimodalities and negative heat capacities. 
Phase transitions have been widely studied in the thermodynamic limit of infinite systems. 
However, in the physical situations considered here (i.e. small systems or non saturating forces), this limit cannot be taken and phase transitions should be reconsidered from a more general point of view. This is for example the case of matter under long range forces like gravitation. Even if these self gravitating systems are very large they cannot be considered as infinite because of the non saturating nature of the force. Other cases are provided by microscopic or mesoscopic systems built out of matter which is known to present phase transitions. Metallic clusters can melt before being vaporized. Quantum fluids may undergo Bose condensation or a super-fluid phase transition. Dense hadronic matter should merge in a quark and gluon plasma phase while nuclei are expected to exhibit a liquid -gas phase transition and a superfluid phase. For all these systems the theoretical and experimental issue is how to define and sign a possible phase transition in a finite system.
In this review we have presented the synthesis of different works which tend to show that phase transitions can be defined as clearly as in the thermodynamic limit. Depending upon the statistical ensemble, i.e. on the experimental situation, one should look for different signals. In the ensemble where the order parameter is free to fluctuate (intensive ensemble), the topology of the event distribution should be studied. A bimodal distribution signals a first-order phase transition. The direction in the observable space in which the distribution is bimodal defines the best order parameter. To survive the thermodynamic limit, the distance between the two distributions, the two "`phases"', 
should scale like the number of particles. This occurrence of a bimodal distribution is equivalent to the alignment of the partition sum zeroes as described by the Yang and Lee theorem. 
In the associated extensive ensemble, the bimodality condition is also equivalent to the requirement of a convexity anomaly in the thermodynamic potential. The first experimental evidences of such a phenomenon have been reported recently in different fields: the melting of sodium clusters \cite{Schmidt},the fragmentation of hydrogen clusters\cite{Farizon}, the pairing in nuclei%
\cite{Melby} and nuclear multifragmentation \cite{Agostino,Tamain,lopezo}.
However, much more experimental and theoretical studies are now expected to progress in this new field of phase transitions in finite systems. Three challenges can thus been assigned to the physics community:
\begin{itemize}
\item  The statistical description of non extensive systems and in particular of open transient finite systems;
\item  The experimental and theoretical study of phase transitions in thoses systems and of the expected abnormal thermodynamics 
\item  The confirmation of the observation of the 
phase transition and the analysis of the associated equation of state properties and the associated phase diagram.
\end{itemize}

\bibliographystyle{aipproc}   


\begin{thebibliography}{99}

\bibitem{Dauxois3}  T. Dauxois et al, 
\textsl{''Dynamics and Thermodynamicsof Systems with Long Range Interactions'' }
Lecture Notes in Physics Vol.602, Springer (2002).


\bibitem{Lynden-Bell2}  D. Lynden-Bell \& R. Wood, Mon. Not. R. Astron. Soc. \textbf{138} (1968)495; D. Lynden-Bell,Physica \textbf{A 263} (1999)293.

\bibitem{lynden-bell}  R. M. Lynden-Bell, Mol. Phys. \textbf{86} (1995) 1353.

\bibitem{ruffo}  J. Barr\'{e}, D. Mukamel and S. Ruffo, Phys. Rev. Lett.
\textbf{87} (2001)030601;
J. Barr\'{e}, F.Bouchet, T.Dauxois and S. Ruffo, J. Stat. Phys. \textbf{119} (2005) 677;
D.Mukamel, S.Ruffo and N. Schreiber, Phys; Rev. Lett. \textbf{95} (2005) 240604.

\bibitem{tatekawa} T.Tatekawa, F.Bouchet, T.Dauxois and S. Ruffo, Phys. Rev. \textbf{E 71} 
			(2005) 056111.
 
\bibitem{Schmidt}  Schmidt M. et al, Phys. Rev. Lett. \textbf{86}{\ (2001)1191}.

\bibitem{Agostino}  {D'Agostino M. et al, Phys. Lett. \textbf{B 473} 219(2000) }.

\bibitem{Farizon}  F. Gobet et al, Phys. Rev. Lett. \textbf{89} (2002) 183403.

\bibitem{Melby}  E. Melby et al, Phys. Rev. Lett. \textbf{83} (1999)3150; 
S.Siem et al, Phys. Rev. \textbf{C65} (2002) 044318.


\bibitem{landau}  L. D. Landau, E. M. Lifshitz, \textsl{''Statistical Physics''}, Pergamon Press (1980) chap.3.

\bibitem{huang}  K.Huang, \textsl{''Statistical Mechanics''}, John Wiley andSons Inc. (1963) chap.5.

\bibitem{tolman}  R. C. Tolman, \textsl{''Principles of statistical mechanics''}, Oxford University Press, London (1962).

\bibitem{rasetti}  M. Rasetti, \textsl{''Modern methods in statistical mechanics''}, World Scientific, Singapore (1986).

\bibitem{jaynes}  E. T. Jaynes, \textsl{''Information theory and statistical mechanics''}, Statistical Physics, Brandeis Lectures, \textbf{vol.3}, 160(1963).

\bibitem{gross}  D. H. E. Gross, \textsl{''Microcanonical thermodynamics:phase transitions in finite systems''}, Lecture Notes in Physics \textbf{vol.66}, Springer (2001).

\bibitem{balian}  R. Balian, \textsl{''From microphysics to macrophysics''},Springer Verlag (1982).

\bibitem{Hill}  T. L. Hill \textsl{''Thermodynamics of small systems''},Dover, New York (1994).

\bibitem{tsallis}  S. Abe and Y. Okamoto, \textsl{''Non extensivestatistical mechanics and its applications''}, Lecture Notes in Physics \textbf{vol.560}, Springer (2001).

\bibitem{chomaz-here} Ph.Chomaz and F.Gulminelli, topical volume "Dynamics and thermodynamics with nuclear degrees of freedom",
  Ph.Chomaz,F.Gulminelli,W.Trautmann,S.Yennello eds., Springer (2006) and references therein.




\bibitem{gross-here} D.H.E.Gross, topical volume "Dynamics and thermodynamics with nuclear degrees of freedom",
  Ph.Chomaz,F.Gulminelli,W.Trautmann,S.Yennello eds., Springer (2006) and references therein.


\bibitem{barre2} F. Bouchet, J. Barr\'e, Journ. Stat. Phys. \textbf{118} (2005) 1073. 

\bibitem{leyvraz}  F. Leyvraz and S. Ruffo, Physica \textbf{A 305} (2002) 58.

\bibitem{Gulminelli}  F. Gulminelli and Ph. Chomaz, Phys. Rev. E \textbf{66}(2002) 046108.

\bibitem{Costeniuc} M.Costeniuc, R.S.Ellis, H.Touchette, J. Math. Phys. \textbf{46} (2005) 063301.

\bibitem{gauss}  M. S. S. Challa and J. H. Hetherington, Phys. Rev. Lett. \textbf{60} (1988) 77;
R.S.Johal, A.Planes, E.Vives, cond-mat/0307646.


\bibitem{Thirring}  W. Thirring, Z. Phys. \textbf{235} (1970) 339.

\bibitem{Huller}  A. Huller, Z. Phys. \textbf{B93} (1994) 401.

\bibitem{Ellis}  R.\ S.\ Ellis, K. Haven and B.\ Turkington J. Stat. Phys. \textbf{101}(2000) 999.

\bibitem{Dauxois}  T. Dauxois, P. Holdsworth and S.\ Ruffo Eur. Phys. J. \textbf{B16}(2000) 659.

\bibitem{Barre}  J.\ Barr\'{e}, D. Mukamel and S.\ Ruffo, Phys. Rev. Lett. \textbf{87} (2001) 030601
and Cond-mat/0209357.  

\bibitem{Ispolatov}  I.\ Ispolatov and E. G. D. Cohen, Physica A \textbf{295}(2001) 475I.


\bibitem{Kastner}  M. Kastner, M. Promberger and A.\ Huller, J. Stat. Phys. \textbf{99} (2000)1251;
M.\ Kastner,J. Stat. Phys. \textbf{109}(2002) 133; M.Kastner, Physica A \textbf{359} (2006) 447.

\bibitem{Gross3}  D. H. E. Gross and E. V. Votyakov, Eur. Phys. J. B \textbf{15} (2000) 115.

\bibitem{Huller2}  A. Huller and M. Pleimling, Int. J. Mod. Phys. C \textbf{13} (2002) 947.

\bibitem{Pleimling}  M. Pleimling,H. Behringer and A. Huller, Phys. Lett. A \textbf{328} (2004) 432;
H.Behringer, J. Stat. Mech. (2005) P06014.

\bibitem{Chomaz02}  Ph.Chomaz and F. Gulminelli, in T. Dauxois et al, Lecture Notes in Physics Vol. 
\textbf{602}, Springer (2002); F.Gulminelli, Ann. Phys. Fr. \textbf{29} (2004) 6.

\bibitem{Chavanis}  P. H. Chavanis and I. Ispolatov, Phys. Rev. E \textbf{66}(2002) 036109.


\bibitem{bose}  C.J.Pethick and H.Smith, \textsl{''Bose Einstein condensation in 
dilute gases''}, Cambridge University Press, Cambridge (2002).

\bibitem{bose2}  A. Minguzzi et al., Phys. Rep. \textbf{395 }(2004) 223.

\bibitem{qgp}  E. V. Shuryak, Phys. Rep. \textbf{391} (2004) 381.

\bibitem{high:energy}  P. Braun-Munzinger et al., Phys.Lett. \textbf{B 596} (2004) 61;
F.Becattini, et al., Phys.Rev. \textbf{C 72} (2005) 064904.
 
\bibitem{clusters}  C. Brechignac et al., Phys. Rev. Lett. \textbf{92} (2004)083401.

\bibitem{Yang}  T. D. Lee and C. N. Yang, Phys. Rev. \textbf{87} (1952) 404.

\bibitem{topology}  Ph. Chomaz, F. Gulminelli and V. Duflot, Phys. Rev. \textbf{E 64} (2001) 046114.

\bibitem{Lee}  K.C. Lee Phys Rev \textbf{E 53} (1996) 6558.

\bibitem{Chomaz3}  Ph. Chomaz, F. Gulminelli Physica \textbf{A 330}(2003) 451.

\bibitem{binder}  K. Binder, D. P. Landau, Phys. Rev. \textbf{B 30} (1984) 1477.

\bibitem{labastie}  P. Labastie and R. L. Whetten, Phys. Rev. Lett. \textbf{65} (1990) 1567.

\bibitem{europhys}  F. Gulminelli, Ph. Chomaz and V.Duflot, Europhys. Lett. \textbf{50} (2000) 434.

\bibitem{DFT}  Ph. Chomaz et al, in preparation


\bibitem{Reinhardt} H. Reinhardt, R. Balian and Y. Alhassid, Nucl. Phys.A413, \textbf{475} (1984).

\bibitem{balian2}  R. Balian, Y. Alhassid, and H. Reinhardt, Phys. Rep. \textbf{131} (1986) 1. 

\bibitem{gauss2} M.Costeniuc, R.S. Ellis, H.Touchette, B.Turkington, Phys. Rev. \textbf{E 73} (2006) 026105.

\bibitem{t:dep:stat}  F. Gulminelli et al, Proc. VI Latin-American Symp. on Nucl. Phys. and Applic., Iguazu, Argentina (2005) to be published in Acta Phys. Hung. A.

\bibitem{Chomaz4}  F.Gulminelli, Ph. Chomaz, Nucl.Phys.A 734 (2004)581; 
               Ph. Chomaz,F. Gulminelli,O. Juillet, Ann. Phys. \textbf{320}(2005) 135.

\bibitem{traps}  C. Menotti, P. Pedri, S. Stringari, Phys. Rev. Lett. %
\textbf{89} (2002) 252402.%

\bibitem{botvina-here}A.Botvina and I.Mishustin, topical volume "Dynamics and thermodynamics with nuclear degrees of freedom",
  Ph.Chomaz,F.Gulminelli,W.Trautmann,S.Yennello eds., Springer (2006) and references therein.

\bibitem{ditoro-here} M.Di Toro, A.Olmi and R.Roy, topical volume "Dynamics and thermodynamics with nuclear degrees of freedom",
  Ph.Chomaz,F.Gulminelli,W.Trautmann,S.Yennello eds., Springer (2006) and references therein.

\bibitem{tamain-here} B.Tamain, topical volume "Dynamics and thermodynamics with nuclear degrees of freedom",
  Ph.Chomaz,F.Gulminelli,W.Trautmann,S.Yennello eds., Springer (2006) and references therein.


\bibitem{Cauley}  J. L. Mc Cauley,\textsl{\ ''Chaos dynamics and fractals''},Cambridge Nonlinear Science Series 2, Cambridge University Press (1993).

\bibitem{Balian3}  R. Balian and M. V\'{e}n\'{e}roni, Phys. Rev. Lett. \textbf{47}(1981)1353; and (1981)1765(E); Ann. of Phys. (N.Y.) \textbf{164}(1985), 334.

\bibitem{Chomaz5}  Ph. Chomaz, Ann. de Phys. (Paris) \textbf{21}(1996) 669.

\bibitem{vanhove} L.Van Hove, Physica \textbf{15} (1949)  951; 
C.N.Yang and T.D.Lee, Phys.Rev. \textbf{87} (1952) 404;
K.Huang, "`Statistical Mechanics"', John Wiley and Sons Inc. (1963), chap.15.2 and appendix C.



\bibitem{ang_mom} 
P.H. Chavanis, M. Rieutord, Astron.Astrophys. \textbf{412} (2003) 1;
D.H.E. Gross, Entropy \textbf{6} (2004) 158; 
A. De Martino, E.V. Votyakov, D.H.E. Gross, Nucl.Phys. \textbf{B 654} (2003) 427.

\bibitem{pettini}  L. Casetti, M. Pettini and E. G. D. Cohen, Phys. Rep. \textbf{337} (2000) 237.

\bibitem{gross-clus}  I. Hidmi, D.H.E. Gross, H.R. Jaqaman, Eur.Phys.J. \textbf {D 20}(2002) 87.

\bibitem{Dauxois2}  T. Dauxois, V. Latora, A. Rapisarda, S. Ruffo \& A.%
Torcini, in Lec. Not. Phys. \textbf{602} Springer (2002).%
 
\bibitem{berry}  T.L.Beck, R.S.Berry, J.Chem.Phys. \textbf{88} (1988) 3910; D.J.Wales and R.S.Berry,
Phys.Rev.Lett. \textbf{73} (1994)2875; 
R.E.Kunz, R.S.Berry, J.Chem.Phys. \textbf{103} (1995) 1904;  
D.J.Wales et al, Adv. Chem.Phys. \textbf{115} (2001) 1; 
R.S.Berry, Israel Journ.Chem. \textbf{44} (2004) 211.

\bibitem{Antonov}  V.A. Antonov Len. Univ.\textbf{\ 7} (1962)135; 
IAU Symp.\textbf{113} (1995) 525.

\bibitem{Hertel}  P. Hertel \& W. Thirring, Ann. of Phys. \textbf{63}(1971) 520.

\bibitem{Chavanis2}  P.H. Chavanis, in Lect. Not in Phys. Vol. \textbf{602}, Springer (2002);
Astron.Astrophys. \textbf{432} (2005) 117.

\bibitem{Padhmanaban}  T. Padhmanaban, in Lect. Not in Phys. Vol. \textbf{602}, Springer (2002).

\bibitem{Katz}  J. Katz, Not.R.Astr.Soc. \textbf{183}(1978) 765.

\bibitem{Promberger}  M. Promberger \& A. Huller, Z. Phys. \textbf{B97}(1995) 341; 
Z. Phys. \textbf{B93}(1994) 401.

\bibitem{Behringer}  H. Behringer, M. Pleimling \& A. H\"{u}ller,
         Journ. Phys. \textbf{A 38} (2005) 973; H. Behringer, J. Phys. \textbf{A 37} (2004) 1443.
 
\bibitem{Kastner3}  M. Kastner \& M Promberger, J. Stat. Phys. \textbf{53}(1988) 795.

\bibitem{Franzosi2}  R.Franzosi, M.Pettini, L.Spinelli, Phys.Rev.\textbf{E 60} (1999) 5009;
                        Phys.Rev.Lett. \textbf{84 }(2000) 2774.%

\bibitem{Franzosi}  R. Franzosi, M. Pettini, Phys.Rev.Lett. \textbf{92} (2004) 60601;
                     and math-ph/0505057, math-ph/0505058.
 
\bibitem{Naudts}  J. Naudts, Cond-mat/0412683 and Europhys.Lett., in press.

\bibitem{Antoni}  M. Antoni, S. Ruffo, A. Torcini, Europhys. Lett., \textbf{66} (2004) 645. 

\bibitem{Gulminelli2}  F. Gulminelli, Ph. Chomaz, A.H. Raduta, A.R. Raduta, Phys. Rev. Lett.                                        \textbf{91}(2003) 202701.

\bibitem{Wynveen}  T.E.Strezelecka, M.W. Davidson, R.L. Rill, Nature \textbf{331} (1988) 457;
Y.Kafri, D.Mukamel, L.Peliti, Phys. Rev. Lett. \textbf{85} (2000) 4988;
A. Wynveen, D.J. Lee \& A.A. Kornyshev, Eur. Phys. J. E \textbf{16} (2005) 303. 

\bibitem{Gulminelli3}  F. Gulminelli, Ph. Chomaz, Phys. Rev. Lett. \textbf{82}(1999) 1402.

\bibitem{Chomaz}  Ph. Chomaz, Gulminelli, Nucl. Phys. \textbf{A 647}(1999) 153.

\bibitem{grossmann}  S.Grossmann and W. Rosenhauer, Z.Phys. \textbf{207} (1967) 138;
P.Borrmann et al., Phys.Rev.Lett.\textbf{84} (2000)3511; 
H.Stamerjohanns et al.,Phys.Rev.Lett.\textbf{88} (2002) 053401.

\bibitem{touchette-comment} H.Touchette, Physica  \textbf{A 359} (2005) 375.

\bibitem{Tamain}  M. Pichon, B. Tamain, R. Bougault,O. Lopez, Nucl. Phys. \textbf{A 749} (2005) 93.

\bibitem{lopezo}  O.Lopez et al., Nucl.Phys. \textbf{A 685} (2001) 246.%

 






\end{thebibliography}




\end{document}